\documentclass[11pt, oneside]{article}   	% use "amsart" instead of "article" for AMSLaTeX format
\usepackage{geometry}                		% See geometry.pdf to learn the layout options. There are lots.
\usepackage{xcolor}
\usepackage{graphicx}				% Use pdf, png, jpg, or eps§ with pdflatex; use eps in DVI mode
\usepackage{amssymb,amsmath,epsfig,cite}

\title{Role of word-of-mouth for programs of voluntary vaccination: A game-theoretic approach}
\author{Samit Bhattacharyya,\footnote{Center for Infectious Disease Dynamics
The Pennsylvania State University, University Park. Email: szb16@psu.edu.}\hskip1.5mm \footnote{to whom the correspondence should be addressed.} \hskip1.5mm Chris Bauch\footnote{Department of Applied Mathematics, University of Waterloo, Waterloo. Email: cbauch@uwaterloo.ca.} \hskip1.5mm and Romulus Breban\footnote{Unit\'e d'Epid\'emiologie des Maladies Emergentes, Institut Pasteur, Paris. Email: romulus.breban@pasteur.fr}}
\date{\today}						

\begin{document}
\maketitle
\bibliographystyle{unsrt}

{\bf Abstract:} We propose a model describing the synergetic feedback between word-of-mouth (WoM) and epidemic dynamics controlled by voluntary vaccination.  We combine a game-theoretic model for the spread of WoM and a compartmental model describing $SIR$ disease dynamics in the presence of a program of voluntary vaccination. We evaluate and compare two scenarios, depending on what WoM disseminates: (1) vaccine advertising, which may occur whether or not an epidemic is ongoing and (2) epidemic status, notably disease prevalence. Understanding the synergy between the two strategies could be particularly important for organizing voluntary vaccination campaigns. We find that, in the initial phase of an epidemic, vaccination uptake is determined more by vaccine advertising than the epidemic status. As the epidemic progresses, epidemic status become increasingly important for vaccination uptake, considerably accelerating vaccination uptake toward a stable vaccination coverage. 

\section{Introduction}
Word-of-mouth (WoM), often fueled by mass media, is a major communication channel for spreading information on vaccines. WoM  disseminates stories on vaccine scares \cite{Bauch:2012ys}, epidemic status \cite{Kalichman:2010zl} and public health interventions \cite{Goldstein:1996qv}; its impact may go both ways. Spread of misperceptions on the relative risks and benefits of vaccination may decrease the vaccination coverage \cite{Chen:2004kq,Samba:2004fj}, while concerns that vaccine would be in short supply may increased compliance with vaccination programs \cite{Goldstein:1996qv}.  

Here we propose a new model describing the synergetic feedback between WoM and epidemic dynamics controlled by voluntary vaccination. WoM may be an important driver of individual-level decisions to vaccinate. In turn, these decisions determine the level of disease immunity in the population and the severity of the epidemic. WoM helps disseminate this information which further motivates voluntary vaccination. We capture these aspects by combining a game-theoretic model for the spread of WoM and a compartmental model describing disease dynamics in the presence of a program of voluntary vaccination.

Previous work focused on modeling the impact of mass media for vaccination programs \cite{Tchuenche:2010eq} and WoM on spreading disease awareness and determining decisions to get vaccinated \cite{Funk21042009,Xia:2014iq}. However, in general, WoM may be charged with mass media soundbites, public health messages and information on epidemic status. The impact of WoM on determining decisions to get vaccinated may be complex. We propose a game-theoretical framework to address these aspects systematically.  In particular, we evaluate and compare two scenarios, depending on what WoM disseminates: (1) vaccine advertising, which may occur whether or not an epidemic is ongoing and (2) epidemic status, notably disease prevalence. We discuss these scenarios starting from an $SIR$ framework. We derive several analytical results regarding $R_0$, vaccination uptake and stable vaccination coverage. We discuss these results from a public health perspective.

\section{Model}
The model consists of two parts. One describes disease transmission and it is modeled using a system of ordinary differential equations. The other describes the spread through WoM (assumed to be much faster than disease transmission) and is modeled using game theory.

\subsection{Framework for Compartmental Model}
\begin{figure}[h!]\begin{center}
\includegraphics[width=0.4\textwidth]{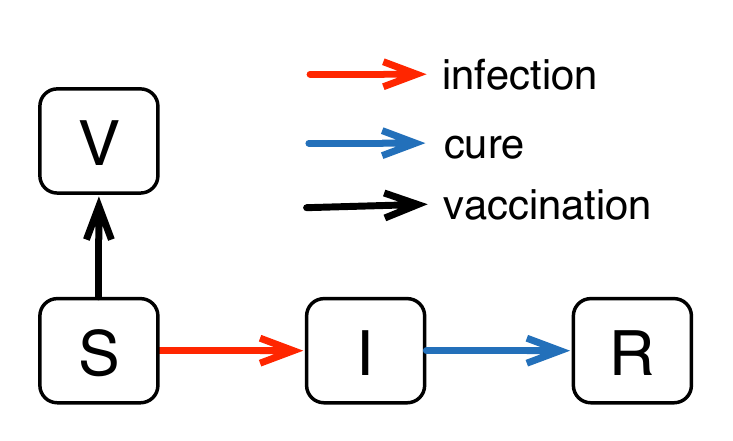}
\caption{Flow diagram of the model of disease natural history where pre-exposure vaccination is offered. The boxes represent population compartments according to the disease status: susceptible $S$, infectious $I$, recovered $R$, and vaccinated $V$. The colored arrows represent the epidemiological processes listed in the figure legend. Demographic processes (i.e., births and disease-unrelated deaths) are not represented in the diagram.}\label{fig:1}\end{center}
\end{figure}

We assume that the public health authority offers pre-exposure vaccination with a perfect vaccine against a disease which is transmitted according to the principles of the $SIR$ model; see Fig.~\ref{fig:1} for the flow diagram.  The vaccine is always available in unrestricted supply, whether or not there is an epidemic.  Vaccination is not enforced; individuals must request to get vaccinated (and show up for vaccination), in order to be vaccinated at no direct cost for themselves.

We propose the following set of differential equations for the model
\begin{eqnarray}
\label{eq:S}
\frac{dS}{dt}&=&\pi-\mu S-\frac{\beta SI}{N}-\mathcal{V},\\
\label{eq:I}
\frac{dI}{dt}&=&\frac{\beta SI}{N}-(\mu+\gamma)I,\\
\label{eq:R}
\frac{dR}{dt}&=&\gamma I-\mu R,\\
\label{eq:V}
\frac{dV}{dt}&=&\mathcal{V}-\mu V,
\end{eqnarray}
where $N=S+I+R+V$. The parameters are as follows. The symbol $\pi$ stands for the birth rate, $\mu$ for disease-unrelated death rate, $\beta$ for disease transmissibility and $\gamma$ for the recovery rate. The term $\mathcal{V}$ represents a function of epidemic variables and parameters, to be determined in the next section. For further convenience, we introduce the symbol $\mathcal{P}$ to denote the prevalence of infectious individuals; i.e., $\mathcal{P}\equiv I/N$.

\subsection{Game theoretical framework}
We assume that individuals meet in pairs and discuss whether or not they got vaccinated and the utility of vaccination. We model this social interaction using the concept of {\it payoff}, adapted from applications of game theory to economics. By payoff we mean an empirical score for the necessity of vaccination that interlocutors reach for themselves, as a result of discussion.  A positive score means that vaccination was necessary, while a negative score means that vaccination would be necessary.  We may also consider that payoff describes the risk of becoming infected and benefit of vaccination.  If the score is positive, then risk is averted and the vaccination is perceived as beneficial. If the score is negative, then risk is present: high risk corresponds to low score while low risk corresponds to high score.  Individuals meet in pairs to {\it play} a coordination game, defined by a {\it payoff matrix} expressing how personal beliefs about vaccination are reinforced; see Table \ref{tab:payoffs}. 
We discuss the interpretation of two payoff matrices, each describing different social phenomena.

\begin{table}[h!]
\begin{minipage}{\linewidth}
\caption{Structure of the payoff matrix for the coordination game of social interaction. Various mathematical expressions functions of epidemiological parameters may be proposed for the entries $a$, $b$, $c$ and $d$.}\vskip2mm
\begin{center}
\begin{tabular}{c|cc}
 	 & $V$ & $S$\\
\hline
	$V$ & $(a, a)$ & $(b,c)$\\	
	$S$ & $(c,b)$ & $(d, d)$ \\
	\end{tabular}
	\end{center}
	\label{tab:payoffs}
	\end{minipage}
\end{table}

\subsubsection{Social interaction driven by vaccine advertising} 
We assume that vaccination benefits are advertised by the health authority in the mass media, while no epidemiological information is accessible through public health. This message could be conveyed even in absence of an epidemic to build herd immunity in the population. We assume that, as a result of advertising alone, all individuals perceive the benefits of vaccination similarly and have scores of magnitude, $a'$.  However, if a person is vaccinated, he considers he did well and his score is positive. If a person is susceptible, he considers he should get vaccinated and the score is negative. Hence, the message in the media gets into WoM. The payoff matrix, denoted by $\boldsymbol M_v$, is given by Eq.~\eqref{eq:Mm}.
\begin{eqnarray}
\label{eq:Mm}
\boldsymbol M_v=\left[\begin{array}{cc}(a',a') & (a',-a')\\ (-a',a') & (-a',-a') \end{array}\right].
\end{eqnarray}

\subsubsection{Social interaction driven by epidemic status}
We consider that the strength of social interaction which spreads through WoM depends on disease prevalence; the payoff matrix, denoted by $\boldsymbol M_e$, is given by Eq.~\eqref{eq:Mw}. We assume that only vaccinated individuals are up to date with the state of the epidemic. Susceptible individuals are oblivious of the epidemic and neither pro nor contra vaccination. When two unvaccinated susceptible individuals discuss, they remain neutral versus vaccination (i.e., they both get zero scores; $d''=0$); this situation serves as reference.  When two vaccinated individuals discuss, their pro-vaccination opinions are reinforced; they both get positive scores, which increase with disease prevalence (i.e., $a''$ is a positive, increasing function of $\mathcal{P}$).  When a susceptible individual discusses with a vaccinated individual, the susceptible individual gets a negative score $c''$ (where $|c''|$ increases with $\mathcal{P}$) and the vaccinated individual a positive score $b''$, increasing with $\mathcal P$
\begin{eqnarray}
\label{eq:Mw}
\boldsymbol M_e=\left[\begin{array}{cc}(a'',a'') & (b'',-|c''|)\\ (-|c''|,b'') & (0,0) \end{array}\right].
\end{eqnarray}

\subsubsection{Score statistics and vaccination coverage dynamics}
\label{sec:scostat}
We consider that individuals mix strongly. The probability of meeting a susceptible is 
\begin{eqnarray}
p_S=\frac{S}{S+V},
\end{eqnarray}
and the probability of meeting a vaccinated individual is
\begin{eqnarray}
p_V=\frac{V}{S+V}.
\end{eqnarray}
We use vaccination scores to model how individuals advise getting vaccinated. We make three assumptions on social behavior. First, susceptible individuals look for advice for two personal questions: ``I am susceptible. Do I need to vaccinate now?" and ``I am susceptible and will vaccinate very soon. Will I regret it?". Second, vaccinated individuals search permanently for confirmation whether their decision to get vaccinated was good or not. Third, an individual provides advice according to the {\it if-I-were-you} approach, disclosing the magnitude of his current score as advice on vaccination.   

After meeting sufficiently many individuals, a susceptible gathers the expected score $|c|p_V+|d|p_S$, while a vaccinated individual gathers the expected score $|a|p_V+ |b|p_S$.   The expected score gathered by an arbitrarily chosen individual as advice to get vaccinated is $(|a|+|c|)p_V+(|b|+ |d|)p_S$. Hence, out of $S$ susceptibles who follow the overall opinion about vaccination,
\begin{eqnarray}
\mathcal{V}\propto S[(|a|+|c|)p_V+(|b|+ |d|)p_S]
\end{eqnarray}
are expected to get vaccinated in the next unit of time. Table \ref{tab:vaccs} provides formulae for the terms $\mathcal{V}$ corresponding to the particular payoff matrices discussed in the previous section. 
\begin{table}[h!]
\begin{minipage}{\linewidth}
\caption{Formulae for the vaccination terms $\mathcal{V}$, given the payoff matrix $\boldsymbol M$.}\vskip2mm
\begin{center}
\begin{tabular}{l|l}
 	 Matrix $\boldsymbol M$& Vaccination term $\mathcal{V}$\\
\hline \hline
	$\boldsymbol M_v$ & $\mathcal{V}_v\propto S(2a')$\\
\hline		
	$\boldsymbol M_e$ & $\mathcal{V}_e\propto S[(a''+|c''|)V+b''S]\slash(V+S)$\\
\hline	
	$\boldsymbol M_v+\boldsymbol M_e$ & $\mathcal{V}_{ve}=\mathcal{V}_v+\mathcal{V}_e$	
	\end{tabular}
	\end{center}
	\label{tab:vaccs}
	\end{minipage}
\end{table}

\section{Results}
We present numerical and analytical results for $\mathcal{V}_{ve}=\mathcal{V}_v+\mathcal{V}_e$, to contrast the role of mass media versus WoM for vaccination coverage.  We introduce further rate parameters to express $\mathcal{V}_{v,e}$. We write $\mathcal{V}_v=\xi S$, just like a {\it classic} vaccination term \cite{1982Sci...215.1053A}. This term is consistent with the public health recommendation to get vaccinated even when disease is not spreading within population. 

\begin{figure}[h!]\begin{center}
\includegraphics[width=0.7\textwidth]{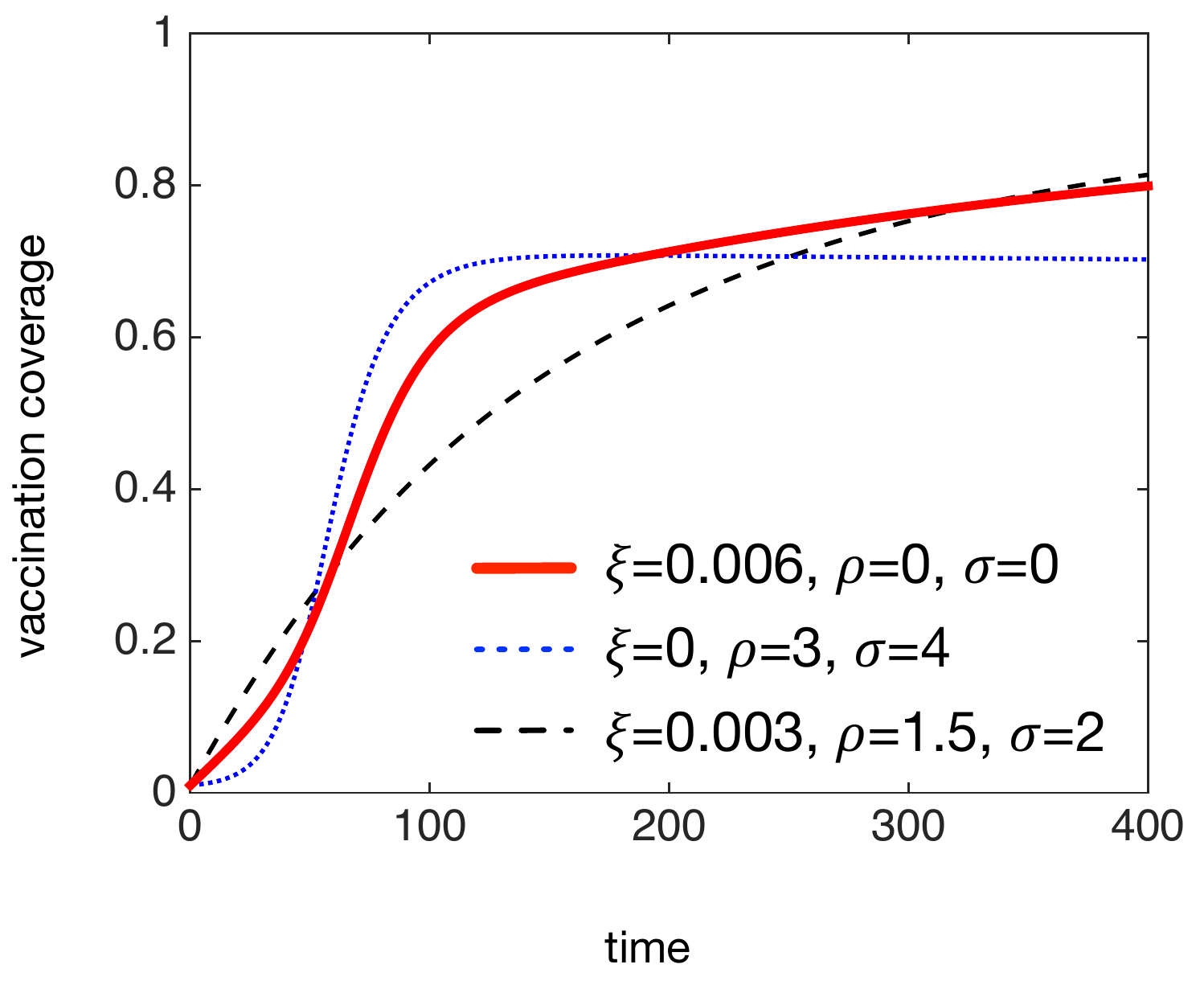}
\caption{Representative graphs of $V(t)$ versus $t$. The parameter values of the model without vaccination are $\pi=0.548$, $\beta=0.2$, $\mu=3.6529\times10^{-5}$ and $\gamma=0.1$. The values for the vaccination parameters, $\xi$, $\rho$ and $\sigma$, are listed in the figure legend. The unit of time is days, so the parameter values correspond to an influenza-like epidemic.}\label{fig:2}\end{center}
\end{figure}

Assuming $a''\propto\mathcal{P}$, $b''\propto\mathcal{P}$ and $c''\propto\mathcal{P}$, we write $\mathcal{V}_e$ in two alternative forms
\begin{eqnarray}
\mathcal{V}_e&=&\left(\frac{SI}{N}\right)\frac{\rho V+\sigma S}{V+S},
\label{eq:Vinc}\\
\mathcal{V}_e&=&S\left[\mathcal{P}\,\left(\frac{\rho V+\sigma S}{V+S}\right)\right].
\label{eq:Vprev}
\end{eqnarray}
Equation \eqref{eq:Vinc} shows that $\mathcal{V}_e$ is positively correlated to disease incidence, while Eq.~\eqref{eq:Vprev} shows that $\mathcal{V}_e$ is positively correlated to the prevalence of infectious individuals. Figure \ref{fig:2} shows graphs of $V(t)$ versus $t$ for various values of the parameters $\xi$, $\rho$ and $\sigma$.

\subsection{The basic reproduction ratio $R_0$}
The disease-free equilibrium ($DFE$) of the model is given by
\begin{eqnarray}
(S_{DFE}, I_{DFE}, R_{DFE}, V_{DFE})=\left(\frac{\pi}{\mu+\xi},0,0,\frac{\pi\xi}{(\mu+\xi)\mu}\right),
\label{eq:dfe}
\end{eqnarray}
independent of $\rho$ and $\sigma$. This equilibrium loses stability as $R_0$, the basic reproduction ratio, exceeds 1. Next generation analysis \cite{VandenDriessche:2002ta} yields the following $R_0$ formula 
\begin{eqnarray}
R_0=\frac{\beta/(1+\xi/\mu)}{(\mu+\nu)}.
\label{eq:r0}
\end{eqnarray}
The parameters $\rho$ and $\sigma$ do not occur in Eq.~\eqref{eq:r0}, as a consequence of the fact that they do not occur in Eqs.~\eqref{eq:I} and \eqref{eq:dfe}. However, $\rho$ and $\sigma$ occur in other key formulae of the model (e.g., the formulae for the endemic state); see \cite{Breban:2014jh} for a similar discussion.

\subsection{Vaccination uptake}
We order the parameters $\xi$, $\rho$ and $\sigma$ depending on their impact on vaccination uptake (i.e., $dV\slash dt$) at disease invasion.  Hence, we consider the model given by Eqs.~\eqref{eq:S}-\eqref{eq:V} under the following assumptions: $N\approx (S+I+V)\sim{\rm const.}$, $S\sim{\rm const.}$, $R=0$, $I/N\ll 1$, and $V/N\ll 1$. Taking partial derivatives of Eq.~\eqref{eq:V} where ${\mathcal V}={\mathcal V}_{ve}$ we obtain
\begin{eqnarray}
\frac{\partial(dV\slash dt)}{\partial\rho}&=&\frac{SIV}{N(V+S)}\sim\frac{IV}{N}\ll I,\\
\frac{\partial(dV\slash dt)}{\partial\sigma}&=&\frac{S^2I}{N(V+S)}\sim I\ll S,\\
\frac{\partial(dV\slash dt)}{\partial\xi}&=&S.
\end{eqnarray}
Hence, the parameter hierarchy for vaccination uptake at disease invasion is (in descending order) $\xi$, $\sigma$ and $\rho$. 
 
\subsection{Time-scales to reach stable vaccination coverage}
The term ${\mathcal V}_v=\xi S$ yields a fixed time scale of $1/\xi$ for susceptible individuals to get vaccinated. The term ${\mathcal V}_e$ yields a time scale between $\min\{\rho, \sigma\}I/N$ and $\max\{\rho, \sigma\}I/N$, which changes over the course of the epidemic. In particular, this time scale decreases as disease prevalence ${\mathcal P}=I/N$ increases, demonstrating that WoM brings susceptible individuals faster to get vaccinated as the epidemic gets worse. 

\subsection{Stable vaccination coverage}
We discuss the relative role of the parameters $\xi$, $\rho$ and $\sigma$ for the stable level of vaccination coverage. We solve for the endemic equilibrium of the model given by Eqs.~\eqref{eq:S}-\eqref{eq:V} where ${\mathcal V}={\mathcal V}_{ve}$. Using star superscript for the variables at the endemic state, we write
\begin{eqnarray}
N^*&=&\pi/\mu,\\
S^*&=&\frac{\pi(\mu+\gamma)}{\beta\mu},\\
I^*&=&\frac{\pi\slash\mu-S^*-V^*}{1+\gamma\slash\mu},\\
R^*&=&\gamma I^*\slash\mu,\\
0&=&(V^*+S^*)(\xi S^*-\mu V^*)+\frac{\mu}{\beta}\left(\frac{\pi}{\mu}-S^*-V^*\right)(\rho V^*+\sigma S^*).
\label{eq:V*}
\end{eqnarray}
We consider the variation of Eq.~\eqref{eq:V*} in $V^*$, $\xi$, $\rho$ and $\sigma$. The constraint between $\delta\xi$, $\delta\rho$ and $\delta\sigma$ when $\delta V^*=0$ yields
\begin{eqnarray}
\delta\xi\frac{\beta}{\mu}=\frac{V^*+S^*-\pi\slash\mu}{V^*+S^*}\left(\delta\rho\frac{V^*}{S^*}+\delta\sigma\right).
\label{eq:delta}
\end{eqnarray}
It is natural to assume that, at the endemic state, $V^*/S^*\sim{\mathcal O}(1)$. Hence, we have that $(V^*+S^*-\pi\slash\mu)\slash(V^*+S^*)\sim{\mathcal O}(1)$, as well. The coefficients of $\delta\xi$, $\delta\rho$ and $\delta\sigma$ could be comparable in Eq.~\eqref{eq:delta}, for maintaining $V^*$ unchanged. That is, changes in $\xi$, $\rho$ and $\sigma$ may not follow a clear hierarchy in determining $V^*/N^*$, the vaccination coverage at the endemic state. 

\section{Discussion}
WoM is an important communication channel that has been previously taken into account in models of sales \cite{Bass:1969ur} and fashion spread \cite{Coulmont:2015a}.  Following previous work on fashion spread \cite{Coulmont:2015a}, we proposed a mixed model which combines a compartmental model for disease transmission with a coordination game for the impact of WoM.  The game analysis provided the vaccination term for the compartmental model, depending on specific assumption on the message and the spread of WoM. Hence, the complete model resulted as a closed system of nonlinear ordinary differential equations amenable to standard analysis.  We used our model to discuss the relative role of WoM for vaccination advertising versus disseminating epidemic status.  

In the case where WoM disseminates vaccine advertising, we obtained that the vaccination term should be simply proportional to the number of susceptibles (i.e., $\xi\neq0$, $\rho=0$, $\sigma=0$). This analytic form has been previously used in compartmental models where vaccination is described as a flow from the susceptible to the recovered class \cite{1982Sci...215.1053A,Yusuf:2012vh}. The flow rate parameter enters the $R_0$ formula, demonstrating that $R_0$ may decrease as a result of vaccination \cite{1982Sci...215.1053A,Yusuf:2012vh}.  Another important result is that such a vaccination strategy yields a disease-free equilibrium for the model that consists from both susceptible and vaccinated populations. In practice, this may cause individuals to get vaccinated even if the community is disease-free. However, keeping the community continuously motivated for getting vaccinated may be difficult and have prohibitive costs.  

A very different scenario would be in place if we considered that WoM disseminated the epidemic status (i.e., $\xi=0$, $\rho\neq0$, $\sigma\neq0$). First, $R_0$ would not change because of vaccination caused by epidemic scares. Second, the disease-free equilibrium would naturally consist only from susceptibles, as individuals would have no incentive to get vaccinated, unless an epidemic occured.

Understanding the synergy between the two strategies could be particularly important for organizing voluntary vaccination campaigns. We found that, in the initial phase of an epidemic, vaccination uptake is determined more by vaccine advertising than the epidemic status. Hence, keeping the community continuously aware of vaccine benefits and motivated to get vaccinated may boost vaccination uptake at disease invasion. However, as the epidemic progresses, epidemic scares become increasingly important for vaccination uptake, considerably accelerating vaccination uptake toward a stable vaccination coverage. Finally, the level of the stable vaccination coverage may be determined by WoM disseminating both vaccine advertising and epidemic scares, to a comparable degree.

In a modern era of tightening resources that limit the advertising budget on communication about vaccines, it can be very useful to harness social processes to disseminate messages of the mass media.  Here, we explored conditions under which WoM could match or exceed mass media effects, and synergistically disseminate different mass media messages.

\end{document}